\documentclass[12pt]{article}
\usepackage{graphicx}
\usepackage[cp1251]{inputenc}  
\usepackage[russian]{babel}
\usepackage{latexsym}
\usepackage{amssymb}
\usepackage{amsmath}

\begin{document}

\makeatletter
\renewcommand*{\@cite}[2]{{#2}}
\renewcommand*{\@biblabel}[1]{#1.\hfill}
\makeatother

\title{Interstellar Extinction in Galactic Cirri in SDSS Stripe 82}
\author{\bf \hspace{-1.3cm}\copyright\, 2022 г. \ \ 
G.A.Gontcharov$^1$\thanks{E-mail: georgegontcharov@yahoo.com},
\and \bf A.V.Mosenkov$^{2,1}$,
\and \bf S.S.Savchenko$^{1,3,4}$,
\and \bf V.B.Il'in$^{1,3,5}$,
\and \bf A.A.Marchuk$^{1,3}$,
\and \bf A.A.Smirnov$^{1,3}$,
\and \bf P.A.Usachev$^{1,3,4}$,
\and \bf D.M.Polyakov$^{1,3}$
\and \bf Z.Shakespear$^{2}$
}
\date{$^1$ Pulkovo Astronomical Observatory, Russian Academy of Sciences, St. Petersburg, 196140 Russia \\
$^2$ Department of Physics and Astronomy, Brigham Young University, N283 ESC, Provo, UT 84602, USA \\
$^3$ St. Petersburg State University, St. Petersburg, 198504 Russia \\
$^4$ Special Astrophysical Observatory, Russian Academy of Sciences, Nizhnii Arkhyz, Karachai-Cherkessian Republic, 369167 Russia \\
$^5$ St. Petersburg State University of Aerospace Instrumentation, St. Petersburg, 190000 Russia}

\maketitle

\newpage

ABSTRACT
We have applied the method of star counts with Wolf diagrams to determine the interstellar extinction in five Galactic cirri in Sloan Digital Sky Survey (SDSS) Stripe 82. 
For this purpose, we have used the photometry of stars in the GALEX NUV filter and the photometry of red dwarfs in five SDSS bands and four SkyMapper Southern Sky Survey DR2 bands. 
We have identified the cirri as sky regions with an enhanced infrared emission from the Schlegel+1998 map. The extinction in them has been calculated relative to the nearby 
comparison regions with a reduced emission. The results for different filters agree well, giving the range of distances and the extinction law for each cirrus. 
The distances in the range 140--415 pc found are consistent with the 3D reddening maps. In the range between the $B$ and $V$ filters the extinctions found are consistent with 
the estimates from Schlegel+1998 for the Cardelli+1989 extinction law with $R_\mathrm{V}=3.1$. However, the extinctions found for all of the filters are best described not by 
the Cardelli+1989 extinction law with some $R_\mathrm{V}=3.1$, but by the inverse proportionality of the extinction and wavelength with its own coefficient for each cirrus. 
In one of the cirri our results suggest a very slight decrease in extinction with wavelength, i.e., a large contribution of gray extinction. 
In the remaining cirri a manifestation of gray extinction is not ruled out either. This is consistent with the previous measurements of the extinction law far from the Galactic midplane.
\bigskip\noindent
\leftline {PACS numbers: 97.10.Zr; 98.35.Pr}
\bigskip\noindent
{\it Keywords:}  Hertzsprung-Russell, color-magnitude, and color-color diagrams; Solar neighborhood in Galaxy; interstellar matter; interstellar dust clouds.
\bigskip

\newpage

\section*{INTRODUCTION}

A century ago Wolf (1923) proposed a method of determining the interstellar extinction in dust clouds from the photometry of samples of stars complete to some magnitude. 
A description of these so-called Wolf diagrams is given, for example, in Parenago (1954, pp. 248, 249) and Kulikovskii (1985, pp. 147--149), while Szomoru and Guhathakurta (1999) and
Gorbikov and Brosch (2010) are among recent studies. The extinction in a cloud is determined with respect to the a fortiori low extinction in a comparison region outside the cloud. 
Extinction makes the stars inside or behind the cloud dimmer than the stars ahead of the cloud or in the comparison region. For this reason, the cumulative (brighter than some
specified magnitude) number of stars of one or more close classes inside or behind the cloud is smaller than that in the comparison region.

When comparing these star counts, the systematic differences in the number of stars of one class for the cloud and the comparison region, for example, at their different distances 
from the Galactic midplane, should be taken into account. This problem is solved either by introducing corrections for the difference in the positions of the cloud and the comparison 
region based on some reliable Galactic model or by examining clouds close to the Sun, i.e., within several hundred parsecs. In addition, when comparing the star counts, 
the unavoidable random fluctuations in the number density of stars and their characteristics should be taken into account. This problem is solved by using a large number (thousands) 
of stars. Consequently, very deep photometric surveys with stars of the most common classes, with a maximum space density, have to be used for the successful application
of Wolf diagrams. In the 20th century the application of Wolf diagrams was made difficult precisely due to the absence of deep surveys and reliable Galactic models as well as 
due to the difficulties in separating the stars of one class from the observed mixture of stars.

Note that Wolf diagrams are usually inapplicable to very dense dust clouds because the gradients in medium characteristics are great in such clouds. In such a situation the 
average cloud characteristics derived from Wolf diagrams become meaningless.

We see the advantages of Wolf diagrams compared to other extinction determination methods in the following.
\begin{enumerate}
\item The sample of stars being used can be a mixture of different classes, provided that one class is dominant.
\item Photometry of low accuracy and only in one filter can be used. As a rule, the accuracy of the result is determined not by the photometric accuracy, but by the distribution of 
the sample in absolute magnitude.
\item The result is precisely the interstellar extinction rather than the reddening, as in many other methods. When using photometry in several filters, this allows the extinction 
in each of them and, consequently, the extinction law, i.e., the wavelength dependence of the extinction, to be determined. Thus, Wolf diagrams are one of the few methods that 
reveal gray or nearly gray extinction, i.e., independent or weakly dependent on wavelength.
\item When using high-quality data, the extinction is determined from Wolf diagrams with a comparatively high accuracy, $\approx0.01$ mag. However, an accompanying shortcoming is
approximately the same accuracy for all filters when using the same sample of stars. Given that the extinction in the ultraviolet (UV) is much greater than that in the visible and even
more so in the infrared (IR), the same accuracy of determining the extinction leads to a drop in the relative accuracy with increasing emission wavelength. This means that Wolf diagrams
are useless for IR photometry: for moderately dense clouds the low extinction found has a very low relative accuracy, while for dense clouds their nonuniform structure makes the
cloud-averaged extinction found meaningless. 
\item Wolf diagrams allow not only the extinction, but also the distance to a cloud to be determined if the distribution of the stars being used in absolute magnitude 
(luminosity function) is known.
\end{enumerate} 

The recent appearance of deep photometric surveys with CCD arrays has revived the application of Wolf diagrams. In fact, their first successful application at a new level was 
presented by Szomoru and Guhathakurta (1999), who also gave an informative overview of the terms and history of this branch of astronomy.

In this study we consider Wolf diagrams for dust clouds in Stripe 82, in which there is deep photometry in many filters. The narrow Stripe 82 of the celestial sphere along the 
celestial equator of width $-1.25^{\circ}<\delta<1.25^{\circ}$, length $-50^{\circ}<\alpha<+60^{\circ}$, and area about 275 deg$^2$ is known primarily for its intensive 
photometric observations in the Sloan Digital Sky Survey (SDSS) $u_\mathrm{SDSS}$, $g_\mathrm{SDSS}$, $r_\mathrm{SDSS}$, $i_\mathrm{SDSS}$, $z_\mathrm{SDSS}$ filters 
(Fliri and Trujillo 2016). The prospects and advantages of the study of clouds at high Galactic latitudes by the Wolf diagram method using SDSS photometry were described by 
Szomoru and Guhathakurta (1999).

In addition to the SDSS photometry, we used the fairly deep photometry of Stripe 82 objects in the NUV (near ultraviolet) filter from the survey performed by the Galaxy Evolution 
Explorer telescope\footnote{https://cdsarc.cds.unistra.fr/viz-bin/cat/II/335} (GALEX; Martin et al. 2005) and presented by Bianchi et al. (2017) as well as in the 
$g_\mathrm{SMSS}$, $r_\mathrm{SMSS}$, $i_\mathrm{SMSS}$, $z_\mathrm{SMSS}$ filters of the SkyMapper Southern Sky Survey DR22\footnote{https://skymapper.anu.edu.au} 
(SMSS; Onken et al. 2019).

Stripe 82 crosses the middle and high Galactic latitudes ($-64^{\circ}<b<-24^{\circ}$), suggesting a comparatively low extinction. Stripe 82 contains filamentary dust clouds of 
small optical depth, i.e., optical and IR high-latitude Galactic cirri (Szomoru and Guhathakurta 1999; Marchuk et al. 2021; and references therein). Our paper is devoted to 
estimating the extinction, the distance, and the extinction law with Wolf diagrams for several Galactic cirri in Stripe 82 using SDSS, SMSS, and GALEX photometry.

\begin{figure*}
\includegraphics{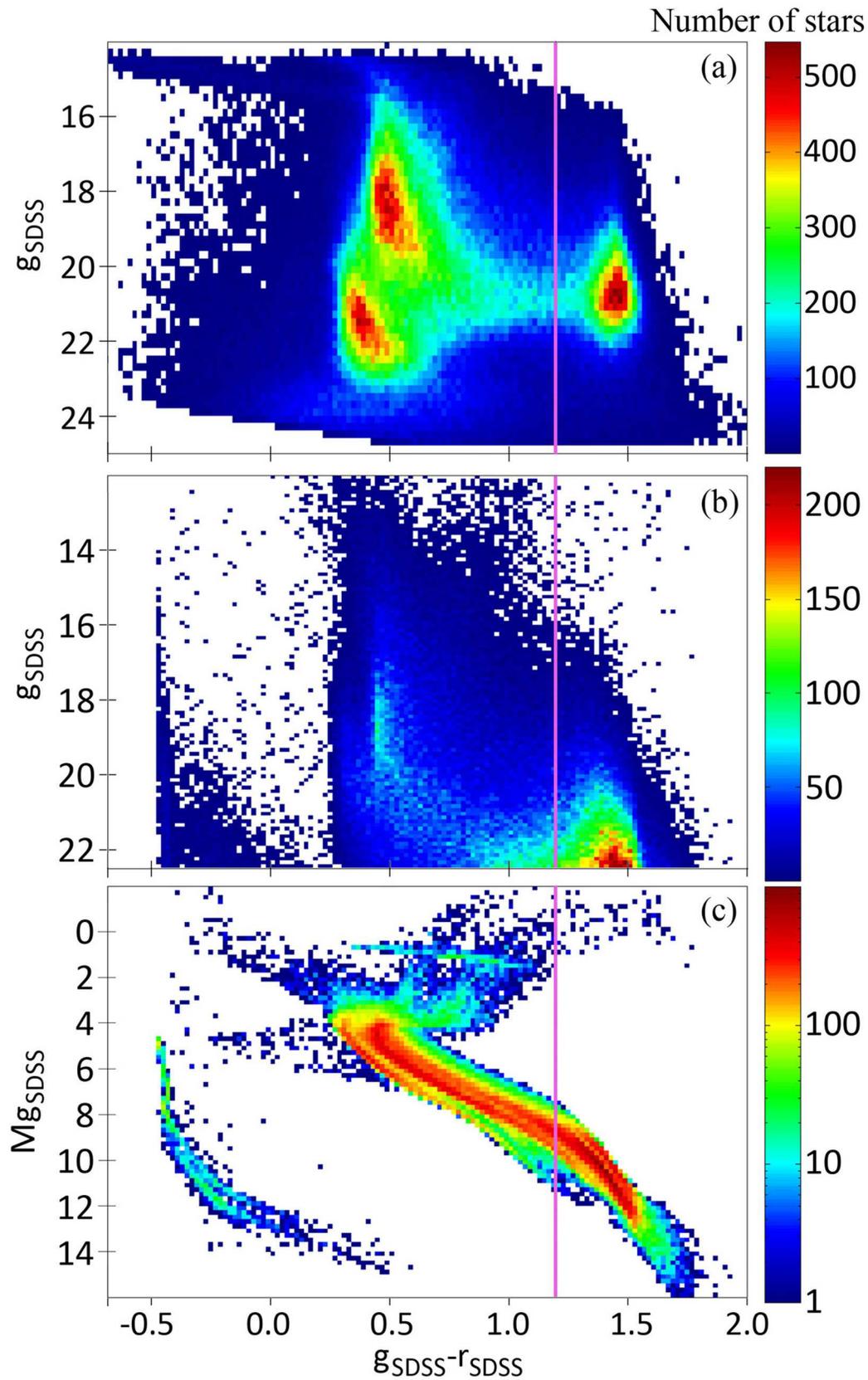}
\caption{(a) The $(g_\mathrm{SDSS}-r_\mathrm{SDSS})$ -- $g_\mathrm{SDSS}$ diagram for the sample of stars from Fliri and Trujillo (2016) in cirrus 1 of Stripe 82,
(b) the same from the TRILEGAL stellar population model, and (c) the corresponding Hertzsprung--Russell $(g_\mathrm{SDSS}-r_\mathrm{SDSS})$ -- $M_\mathrm{g_\mathrm{SDSS}}$ 
diagram from TRILEGAL. The vertical straight line separates the red dwarfs with $g_\mathrm{SDSS}-r_\mathrm{SDSS}>1.2$ used by us.
}
\label{area1}
\end{figure*}

\begin{table*}
\def\baselinestretch{1}\normalsize\normalsize
\caption[]{The cirrus -- comparison region extinction difference $\Delta A$ found for various cirri and photometric filters.
$\lambda_\mathrm{eff}$ is the effective wavelength of the filter in nanometers.
}
\label{solution}
\[
\begin{tabular}{lrccccc}
\hline
\noalign{\smallskip}
Filter  & $\lambda_\mathrm{eff}$ & Cirrus 1 & Cirrus 2 & Cirrus 3 & Cirrus 4 & Cirrus 5 \\
\hline
\noalign{\smallskip}
GALEX NUV &  231           & $0.232\pm0.04$ & $0.401\pm0.04$ & $0.098\pm0.04$ & $0.194\pm0.04$ & $0.298\pm0.04$ \\ 
$u_\mathrm{SDSS}$  &  353  & $0.197\pm0.03$ & $0.196\pm0.03$ & $0.094\pm0.03$ & $0.119\pm0.03$ & $0.207\pm0.03$ \\
$g_\mathrm{SDSS}$  &  475  & $0.194\pm0.03$ & $0.260\pm0.03$ & $0.082\pm0.03$ & $0.059\pm0.03$ & $0.173\pm0.03$ \\
$r_\mathrm{SDSS}$  &  622  & $0.165\pm0.03$ & $0.207\pm0.03$ & $0.058\pm0.03$ & $0.029\pm0.03$ & $0.129\pm0.03$ \\
$i_\mathrm{SDSS}$  &  763  & $0.197\pm0.03$ & $0.149\pm0.03$ & $0.050\pm0.03$ & $0.033\pm0.03$ & $0.125\pm0.03$ \\
$z_\mathrm{SDSS}$  &  905  & $0.192\pm0.03$ & $0.100\pm0.03$ & $0.053\pm0.03$ & $0.019\pm0.03$ & $0.119\pm0.03$ \\
$g_\mathrm{SMSS}$  &  514  & $0.175\pm0.03$ & $0.193\pm0.03$ & $0.096\pm0.03$ & $0.038\pm0.03$ & $0.129\pm0.03$ \\
$r_\mathrm{SMSS}$  &  615  & $0.177\pm0.03$ & $0.144\pm0.03$ & $0.072\pm0.03$ & $0.044\pm0.03$ & $0.116\pm0.03$ \\
$i_\mathrm{SMSS}$  &  776  & $0.169\pm0.03$ & $0.143\pm0.03$ & $0.089\pm0.03$ & $0.041\pm0.03$ & $0.110\pm0.03$ \\
$z_\mathrm{SMSS}$  &  913  & $0.188\pm0.03$ & $0.096\pm0.03$ & $0.084\pm0.03$ & $0.050\pm0.03$ & $0.075\pm0.03$ \\
\hline
\end{tabular}
\]
\end{table*}

\section*{DATA}

In the SDSS\footnote{https://cdsarc.cds.unistra.fr/viz-bin/cat/J/MNRAS/456/1359} data we left only the stars, but not the galaxies (in the data of Fliri and Trujillo (2016) 
the star/galaxy separation is $S/G=1$, while the object image Ellipticity$\le0.4$ suggests that the image is not too elongated) with a good image quality (Flags$\le10$). 
We also constrained the photometric accuracy: higher than 0.4 mag in the $u_\mathrm{SDSS}$ filter and 0.25 mag in the $g_\mathrm{SDSS}$, $r_\mathrm{SDSS}$ and $i_\mathrm{SDSS}$ filters. 
Note that for moderately faint stars the SDSS photometry is comparatively accurate: for example, these constraints on the photometric accuracy eliminate less than 1\% of the stars 
with $g_\mathrm{SDSS}<24$.

The distribution of the selected stars on the color--magnitude `$(g_\mathrm{SDSS}-r_\mathrm{SDSS})$ -- $g_\mathrm{SDSS}$' diagram in a part of Stripe 82 (cirrus 1 discussed below) 
is shown in Fig. 1a. In other parts of Stripe 82 this distribution is analogous. Three regions with an enhanced number density of stars on the color--magnitude diagram are
seen in Fig. 1a, suggesting the dominance of three classes of stars in the sample.

We determined the characteristics of these classes by simulating the stellar population of the sky regions under consideration with the 
TRILEGAL\footnote{http://stev.oapd.inaf.it/cgi-bin/trilegal} model (Girardi et al. 2005) by taking into account the PARSEC\footnote{http://stev.oapd.inaf.it/cgi-bin/cmd\_3.6} 
stellar structure and evolution model (Bressan et al. 2012). For comparison with the observed diagram, Fig. 1b shows the model color--magnitude diagram, while Fig. 1c shows 
the corresponding model Hertzsprung--Russell (HR) diagram for cirrus 1 under the constraint $g_\mathrm{SDSS}<22.5$ that is close to the actual constraint in Fig. 1a. 
We adopted the TRILEGAL parameters by default, but with more realistic extinction estimates for the cirri and the comparison regions. It can be seen that dwarfs of
three classes dominate among the stars:
\begin{itemize}
\item the G dwarfs with an approximately solar metallicity located near the main-sequence (MS) turnoff and belonging to the thin disk form a clump of comparatively bright and
hot stars on the color--magnitude diagram in Figs. 1a and 1b at $g_\mathrm{SDSS}-r_\mathrm{SDSS}\approx0.55$ and $g_\mathrm{SDSS}\approx18$;
\item the G subdwarfs with a low metallicity that are also located near the MS turnoff, but belong to the thick disk form a second MS on the HR diagram below and leftward of the
main one and a clump of comparatively dim and hot stars on the color--magnitude diagram at $g_\mathrm{SDSS}-r_\mathrm{SDSS}\approx0.4$ and $g_\mathrm{SDSS}\approx21.5$ in
Figs. 1a and 1b, with this clump being barely noticeable in Fig. 1b, since TRILEGAL apparently underestimates the number of thick-disk subdwarfs toward cirrus 1;
\item the M red dwarfs with various metallicities predominantly in the range $M/H=-0.25\pm0.30$ form a clump of cool stars on the color--magnitude diagram in Figs. 1a and 1b.
\end{itemize}
The K dwarfs and subdwarfs in Figs. 1a and 1b are much fewer because they are vastly inferior to the M class in number and to the G class in luminosity.

Note that the TRILEGAL simulation results and, accordingly, the color--magnitude and HR diagrams vary noticeably, depending on the adopted initial mass function and binary star 
inclusion parameters. In any case, however, TRILEGAL shows the dominance of the three classes of dwarfs noted by us.

As noted previously, using the stars of only one class increases the accuracy of the Wolf diagrams. Therefore, we decided to use only the M red dwarfs selected with the constraint
$g_\mathrm{SDSS}-r_\mathrm{SDSS}>1.2$, as shown in Fig. 1. A reasonable variation of this selection criterion changes the sample composition by less than 2\%, i.e., insignificantly. 
Apart from other advantages, using red dwarfs must reduce sharply the admixture of unidentified galaxies in the sample, since the overwhelming majority of galaxies observed in the
SDSS project are much bluer than the red dwarfs at a low redshift and much fainter than the stars of our sample at a high redshift, as shown by Strateva et al. (2001) and 
Fliri and Trujillo (2016).

To determine the median characteristics of our sample of red dwarfs, note that in Fig. 1 the maximum of their color distribution occurs at an observed color 
$g_\mathrm{SDSS}-r_\mathrm{SDSS}\approx1.48$. For this sky region the popular reddening map from Schlegel et al. (1998, hereafter SFD98) gives $E(B-V)\approx0.1$, i.e., 
$E(g_\mathrm{SDSS}-r_\mathrm{SDSS})\approx0.1$ when using the most popular extinction law from Cardelli et al. (1989, hereafter CCM89) with an extinction-to-reddening ratio 
$R_\mathrm{V}\equiv A_\mathrm{V}/E(B-V)=3.1$. Then, the median of the dereddened color of the red dwarfs being used is $(g_\mathrm{SDSS}-r_\mathrm{SDSS})_0\approx1.38$, 
the corresponding median of the mass is about 0.48 solar mass, the median of the absolute magnitude is $M_\mathrm{g_\mathrm{SDSS}}\approx10.7$, and the median of the metallicity is
[M/H]$=-0.2$, in agreement with TRILEGAL and PARSEC.

Let us estimate how far the Galactic cirri that can be investigated using the sample of red dwarfs under consideration can be. Fliri and Trujillo (2016) showed that their separation 
of the observed objects into stars and galaxies is reliable at $g_\mathrm{SDSS}<23$. Therefore, we used fainter objects only to check the results. It can be seen from Fig. 1 that 
the number of red dwarfs decreases at $g_\mathrm{SDSS}>21.5$. However, this decrease is caused no so much by the sample incompleteness as by the significant decrease in the space 
density of red dwarfs with increasing distance from the Galactic plane, since high Galactic latitudes are considered. Since the overwhelming majority of red dwarfs used
by us have $M_\mathrm{g_\mathrm{SDSS}}<13$ and an extinction $A_\mathrm{g_\mathrm{SDSS}}<0.5$ (as follows from TRILEGAL, PARSEC, and the results of our study), the sample 
completeness to $g_\mathrm{SDSS}<23$ is enough to analyze the clouds at a distance up to $R=10^{(23+5-13-0.5)/5}\approx800$ pc from the Sun. This estimate is confirmed by our 
simulations with TRILEGAL. The most reliable distance estimates for the cirri in Stripe 82 under consideration are given by the 3D reddening maps of 
Gontcharov (2017a, hereafter G17),\footnote{https://cdsarc.cds.unistra.fr/viz-bin/cat/J/PAZh/43/521} 
Green et al. (2019, hereafter GSZ19),\footnote{http://argonaut.skymaps.info/} and 
Lallement et al. (2022, hereafter LVB22).\footnote{https://cdsarc.cds.unistra.fr/viz-bin/cat/J/A+A/661/A147} 
The first, second, and third maps place the cirri under consideration at distances of 100--320, 140--380, and 120--380 pc from the Sun, respectively.
Thus, the Galactic cirri so close to us may well be analyzed with the sample of red dwarfs under consideration.

In the SMSS data we left the objects with high-quality photometry (all Flags$=0$). As from SDSS, we selected only red dwarfs from SMSS and reduced significantly the admixture of 
galaxies with the constraint $g_\mathrm{SMSS}-r_\mathrm{SMSS}>0.7$.\footnote{This gives qualitatively the same separation as does the criterion $g_\mathrm{SDSS}-r_\mathrm{SDSS}>1.2$. 
The difference in the numbers is caused by the difference between the gSDSS and gSMSS filters.} 
The SMSS photometry has a high accuracy. We eliminated only about 0.01\% of the red dwarfs that demonstrated a photometric accuracy less than 0.3 mag at least in one of the SMSS filters.

To eliminate the galaxies, in the GALEX data we left only the objects that are stars with a probability $>50\%$ and, at the same time, marked as not an extended object. 
Note that these constraints eliminate only 3\% of the objects with GALEX NUV photometry in Stripe 82 and do not affect the results. At the same time, we did not constrain the data
by the accuracy of the GALEX NUV photometry. This accuracy is rather low: its median is 0.4 mag. However, as noted previously, the accuracy of the derived extinction is determined 
not by the photometric accuracy, but by the standard deviation of the absolute magnitude of the stars under consideration, which is greater than one magnitude, according to
PARSEC and TRILEGAL. The stars with GALEX NUV photometry are bright in the UV and include OB stars, blue subdwarfs, horizontal-branch giants with a low metallicity, and other classes. 
According to PARSEC and TRILEGAL, they all span approximately the same range of absolute magnitudes. Therefore, we did not separate any class of stars from the GALEX objects. 
This and the smaller number of stars with GALEX NUV photometry than that with photometry in the optical SDSS and SMSS filters led to a slightly lower accuracy of determining the
extinction in the GALEX NUV filter, as can be seen from Table 1 with our results.

For the application of Wolf diagrams the sky regions with a presumed enhanced extinction must be separated from the comparison regions with a reduced extinction. For this separation 
we used the estimates of the reddening $E(B-V)$  along the line of sight to infinity from SFD98. This approach in separating the cloud and the comparison region to analyze the 
Wolf diagrams was previously applied, for example, by Szomoru and Guhathakurta (1999). The reddening estimates were obtained in SFD98 by calibrating the far-IR dust emission 
estimates based on the color of elliptical galaxies. In turn, these emission estimates were obtained by the Cosmic Background Explorer (COBE) and the Infrared Astronomical 
Satellite (IRAS). Note that the reddening estimates in Stripe 82 based on other studies are very close to the SFD98 estimates. For example, the reddening estimates from the 
Planck Space Telescope observations (2014) calibrated by Meisner and Finkbeiner (2015) based on the color of SDSS stars virtually coincide with the SFD98 estimates: their
correlation coefficient is 0.97.

\begin{figure*}
\includegraphics{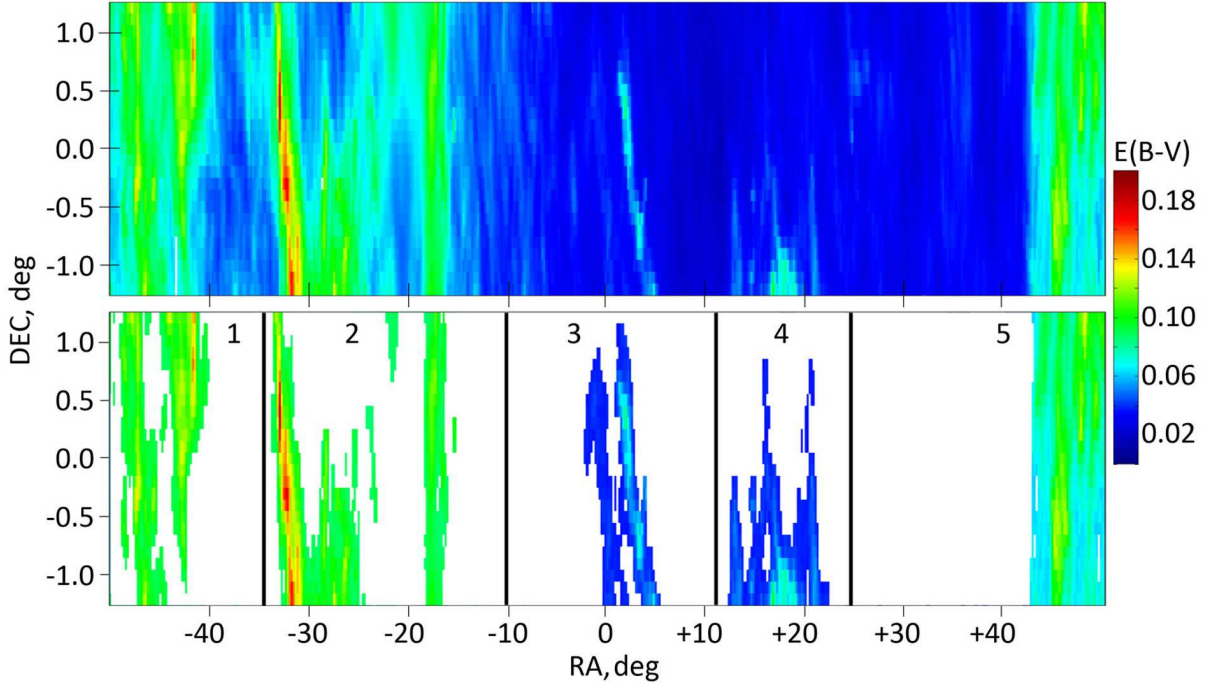}
\caption{Top: reddening $E(B-V)$ from the SFD98 map in the part of Stripe 82 under consideration. Bottom: the Stripe 82 regions attributed to the cirri under consideration are 
marked by the colors and numbers. The vertical lines approximately demarcate the sky regions including the cirrus, its comparison region, and the region that belongs neither to 
the cirrus nor to the comparison region.
}
\label{stripe82}
\end{figure*}

\begin{table*}
\def\baselinestretch{1}\normalsize\normalsize
\caption[]{Characteristics of the cirri under consideration.
Approximate Galactic coordinates $l$ and $b$ of the centers of the cirri under consideration, the cirrus area in deg$^2$ adopted by us, the adopted constraint on the SFD98 reddening 
for the cirrus $E(B-V)_\mathrm{cirrus}$, the median of the SFD98 reddening for the cirrus, the adopted constraint on the SFD98 reddening for the comparison region $E(B-V)_\mathrm{bg}$, 
the median of the SFD98 reddening for the comparison region, the difference of the reddening medians $\Delta E(B-V)$ for the cirrus and the corresponding comparison region, the
extinction estimate $\Delta A_\mathrm{V\,SFD98}$ for the cirrus based on $\Delta E(B-V)$ and the CCM89 extinction law with $R_\mathrm{V}=3.1$.
}
\label{cirrus}
\[
\begin{tabular}{lccccc}
\hline
\noalign{\smallskip}
Characteristic  & Cirrus 1 & Cirrus 2 & Cirrus 3 & Cirrus 4 & Cirrus 5 \\
\hline
\noalign{\smallskip}
$l$, deg                              & $49^{\circ}$  &  $63^{\circ}$  &  $101^{\circ}$ &  $133^{\circ}$ &  $178^{\circ}$ \\ 
$b$, deg                              & $-28^{\circ}$ &  $-45^{\circ}$ &  $-62^{\circ}$ &  $-63^{\circ}$ &  $-48^{\circ}$ \\
Area, deg$^2$                          & $13.9^{\Box}$ & $18.2^{\Box}$  & $8.7^{\Box}$   & $9.5^{\Box}$   & $18.2^{\Box}$  \\
$E(B-V)_\mathrm{cirrus}$         & $>0.088$      & $>0.085$       & $>0.035$       & $>0.035$       & $>0.060$      \\
Median $E(B-V)_\mathrm{cirrus}$ & 0.100         & 0.103          & 0.039          & 0.040          & 0.081    \\
$E(B-V)_\mathrm{bg}$             & $<0.064$      & $<0.050$       & $<0.021$       & $<0.021$       & $<0.032$   \\
Median $E(B-V)_\mathrm{bg}$     & 0.052         & 0.046          & 0.019          & 0.020          & 0.028    \\
$\Delta E(B-V)$                  & 0.048         & 0.057          & 0.020          & 0.020          & 0.053    \\
$\Delta A_\mathrm{V\,SFD98}$     & 0.148         & 0.176          & 0.062          & 0.062          & 0.163    \\
\hline
\end{tabular}
\]
\end{table*}

Figure 2 shows the variations in reddening $E(B-V)$ in the part of Stripe 82 with $-50^{\circ}<\alpha<+50^{\circ}$ under consideration. We do not consider the small part of Stripe 82 
with $+50^{\circ}<\alpha<+60^{\circ}$, since, according to all estimates, the extinction in it is too high ($A_\mathrm{V}>1$) to obtain accurate results with Wolf diagrams. 
Note that the pixelization in Fig. 2 reflects the angular resolution of 6.1 arcmin for the SFD98 map.

Table 2 gives approximate Galactic coordinates of the centers of the cirri under consideration, their areas, and other characteristics. As the cirri we consider the isolated sky 
regions with a reddening above the limit $E(B-V)_\mathrm{cirrus}$ specified in Table 2. Each cirrus has its own limit. The five cirri under consideration are shown in the lower half 
of Fig. 2. Similarly, as the comparison regions we consider the neighborhoods of the cirri with a reddening below the limit $E(B-V)_\mathrm{bg}$ specified in Table 2. The comparison
regions for different cirri can overlap. An important characteristic of each cirrus or comparison region is the reddening median specified in Table 2. We consider the difference of 
the reddening medians for the cirrus and its comparison region $\Delta E(B-V)$ as a reddening estimate basically inside the cirrus. For the study to be effective, $\Delta E(B-V)$ 
must be high enough and, consequently, apart from the cirri and the comparison regions, there are vast regions in Stripe 82 that we attributed to neither of them. In addition, 
the limiting values of $E(B-V)_\mathrm{cirrus}$ and $E(B-V)_\mathrm{bg}$ are chosen so that the areas of the cirrus and its comparison region are exactly equal and large
enough for theWolf diagrams to be applicable.

For each cirrus our estimates of the extinction in it can be compared with the estimate of the extinction $\Delta A_\mathrm{V\,SFD98}$ from Table 2, which was calculated based on 
$\Delta E(B-V)$ from the SFD98 data. One should not expect a close coincidence of our extinction estimates with those from SFD98, since the estimation methods and the initial 
data are completely different. In addition, the SFD98 estimates completely disregard the possible significant variations of the extinction law at high latitudes detected by
Gorbikov and Brosch (2010), Gontcharov (2012), and other authors, whereas the Wolf diagrams are able to show such variations. Moreover, as claimed by the authors of SFD98 themselves 
and shown by many further studies, the estimates of $E(B-V)$ from SFD98 contain significant random and systematic errors (see Gontcharov (2016b) for a review and 
Gontcharov and Mosenkov (2017, 2018, 2021) for a discussion). For example, the uncertainty in the IR emission–reddening calibration declared by the authors of SFD98 is 
$\sigma(E(B-V))=0.028$. On the other hand, the errors of the extinctions found by us are also significant. Therefore, the comparison of our results with SFD98 presented in the Section
`Results' is very interesting.

Table 2 shows that for the sake of a sufficiently large number of stars in the cirri and the comparison regions we are forced to increase their area and, consequently, to include 
regions with comparatively low and high reddenings, respectively, in them. As a result, the medians of $E(B-V)_\mathrm{bg}$ are rather large, especially in cirri 1 and 2. 
Cirri 3 and 4 have the smallest $\Delta E(B-V)$ (i.e., do not differ greatly from the background) and the smallest area. This makes the results for them least reliable.

Stripe 82 is too narrow to completely accommodate the cirri crossing it. However, they are clearly seen completely on the SFD98 map. It shows that although cirri 1, 2, and 5 
inside Stripe 82 can be separated into individual filaments, outside Stripe 82 the filaments attributed by us to one cirrus do join together. To justify our consideration of 
cirri 1, 2, and 5 as whole objects, we calculated the extinctions and distances using Wolf diagrams not only for the whole cirri, but also for their parts visually distinguishable
in Fig. 2 and obtained results coincident with one another within their uncertainty limits.

\begin{figure*}
\includegraphics{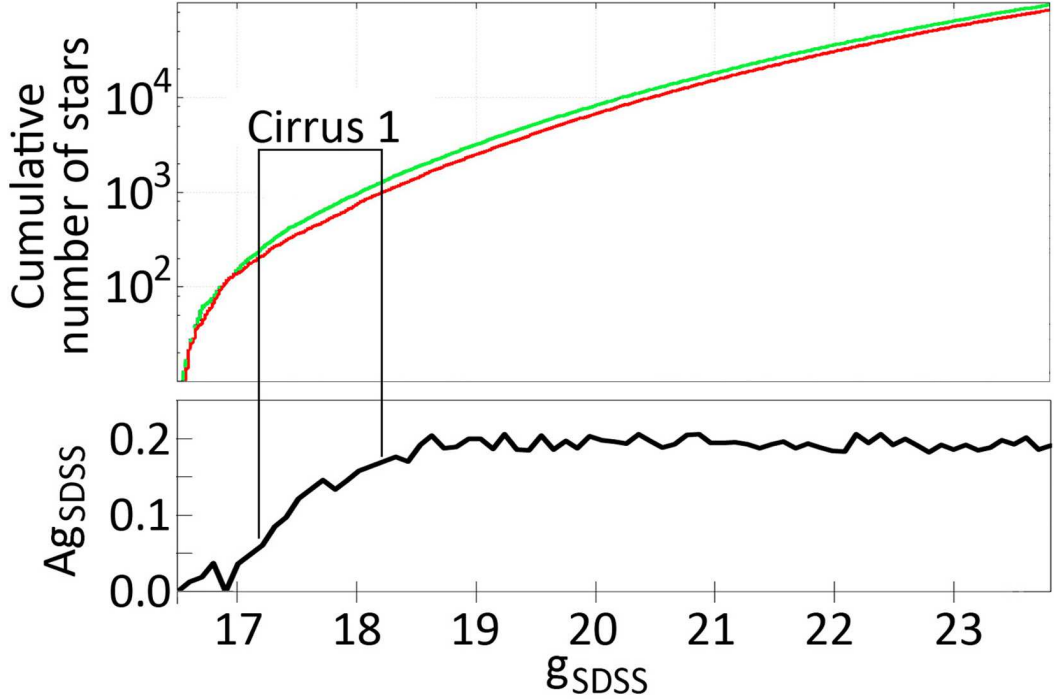}
\caption{Wolf diagramfor cirrus 1 and the $g_\mathrm{SDSS}$ filter. Top: the cumulative number of red dwarfs as a function of $g_\mathrm{SDSS}$ in regions with small and large IR 
dust emission -- the upper (green) and lower (red) curves, respectively. Bottom: the difference between these curves along the horizontal axis is the extinction 
$A_\mathrm{g_\mathrm{SDSS}}$ as a function of $g_\mathrm{SDSS}$. The range of a significant increase in $A_\mathrm{g_\mathrm{SDSS}}$ is marked as cirrus 1.
}
\label{rdarea1g}
\end{figure*}

\begin{table*}
\def\baselinestretch{1}\normalsize\normalsize
\caption[]{The range $R$ of distances to the cirri found (parsecs). 
The found range $R$ is the range of distances to the cirri found; $R_\mathrm{G17}$, $R_\mathrm{GSZ19}$, and $R_\mathrm{LVB22}$ are the estimates of the range of distances to the 
cirrus from the maps of Gontcharov (2017a), Green et al. (2019), and Lallement et al. (2022), respectively. The $R^{*}$ range of overlap of the estimates is specified in the lower row. 
For cirrus~4 $R^{*}$ is specified without taking into account the estimate from LVB22.
}
\label{distance}
\[
\begin{tabular}{lccccc}
\hline
\noalign{\smallskip}
Distance, pc     & Cirrus 1 & Cirrus 2 & Cirrus 3 & Cirrus 4 & Cirrus 5 \\
\hline
\noalign{\smallskip}
$R$                & 150--310 & 195--415  & 150--270 & 170--300 & 140--340 \\
$R_\mathrm{G17}$   & 120--200 & 100--280  & 160--320 & 160--200 & 140--200 \\
$R_\mathrm{GSZ19}$ & 200--380 & 140--210  & 170--230 & 170--190 & 150--200 \\
$R_\mathrm{LVB22}$ & 120--260 & 180--340  & 230--380 & 270--360 & 130--210 \\
$R^{*}$            &  200     & 195--210  &   230    & 170--190 & 150--200 \\
\hline
\end{tabular}
\]
\end{table*}

\begin{figure}
\includegraphics{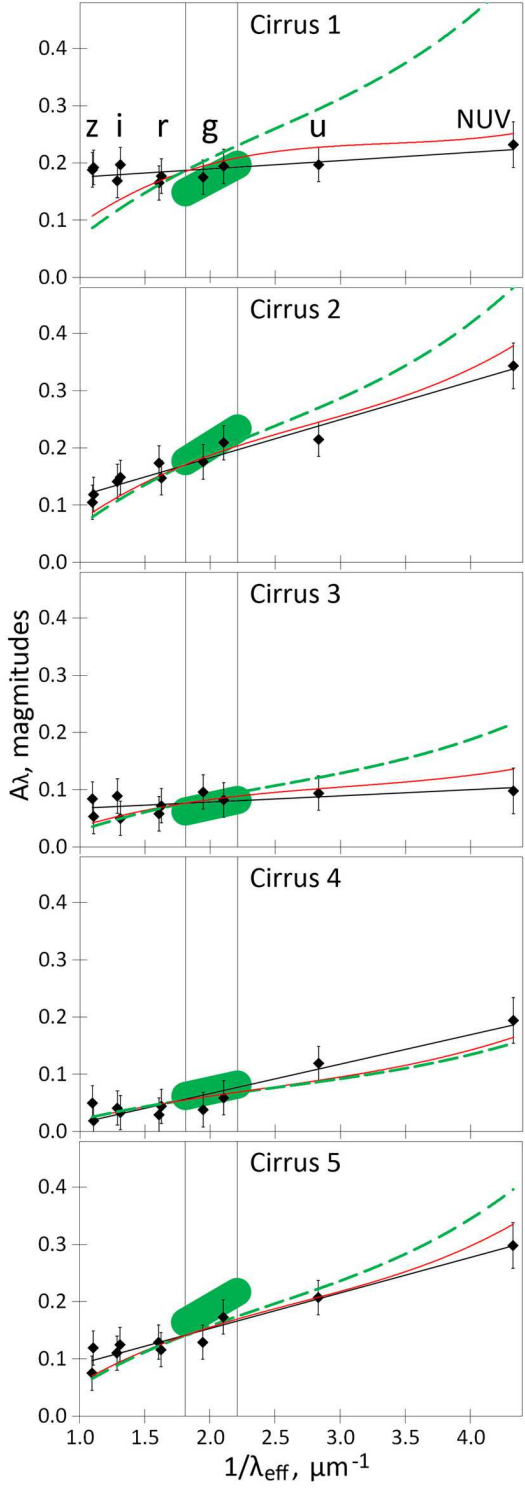}
\caption{Extinction $A_\mathrm{\lambda_\mathrm{eff}}$ versus reciprocal of the wavelength $\lambda_\mathrm{eff}$ in micrometers for the five cirri under consideration: the
diamonds represent the extinctions found, the black straight line indicates their linear fit, the thick green dashed line indicates the fit by the CCM89 extinction law with 
$R_\mathrm{V}=3.1$, the thin red curve indicates the fit by the CCM89 extinction law with the optimal $R_\mathrm{V}$, the vertical straight lines mark the effective wavelengths 
of the $B$ and $V$ filters, the green region is the prediction based on SFD98 with the CCM89 extinction law at $R_\mathrm{V}=3.1$ including the uncertainty. 
The letters mark the pairs of $z_\mathrm{SMSS}$ and $z_\mathrm{SDSS}$, $i_\mathrm{SMSS}$ and $i_\mathrm{SDSS}$, $r_\mathrm{SDSS}$ and $r_\mathrm{SMSS}$,
$g_\mathrm{SMSS}$ and $g_\mathrm{SDSS}$ filters, and the $u_\mathrm{SMSS}$ and GALEX NUV filters.
}
\label{law}
\end{figure}

\begin{table*}
\def\baselinestretch{1}\normalsize\normalsize
\caption[]{The median of the $g_\mathrm{SDSS}-z_\mathrm{SDSS}$ color for the selected galaxies in the cirri and their comparison regions.
The standard deviation of the mean is specified as the uncertainty.
}
\label{gz}
\[
\begin{tabular}{lcc}
\hline
\noalign{\smallskip}
          & Cirrus & Comparison region \\
\hline
\noalign{\smallskip}
Cirrus 1 & $1.33\pm0.01$ & $1.38\pm0.01$ \\
Cirrus 2 & $1.40\pm0.01$ & $1.39\pm0.01$ \\
Cirrus 3 & $1.39\pm0.01$ & $1.40\pm0.01$ \\
Cirrus 4 & $1.39\pm0.01$ & $1.41\pm0.01$ \\
Cirrus 5 & $1.38\pm0.01$ & $1.38\pm0.01$ \\
\hline
\end{tabular}
\]
\end{table*}

\section*{RESULTS}

A typical Wolf diagram for cirrus 1 and the $g_\mathrm{SDSS}$ filter is shown in Fig. 3. Several tens of thousands of red dwarfs were used for it. The measured extinction
$A_\mathrm{g_\mathrm{SDSS}}$ corresponds to the distance along the horizontal axis between the cumulative numbers of red dwarfs in the comparison region and the corresponding cirrus,
i.e., between the color curves. The extinction $A_\mathrm{g_\mathrm{SDSS}}$ is presented on the lower panel in Fig. 3. It can be seen that this extinction grows in the magnitude
range $17.2<g_\mathrm{SDSS}<18.2$ and then is stabilized at $A_\mathrm{g_\mathrm{SDSS}}=0.194$. This magnitude range corresponds to the distance range $186<R<270$ pc for cirrus 1
if we adopt the median absolute magnitude $M_\mathrm{g_\mathrm{SDSS}}=10.7$ for the red dwarfs being used, the background extinction ahead of the cirrus 
$A_\mathrm{g_\mathrm{SDSS}}\approx0.15$ (judging by the data on reddening in the comparison region), and the extinction inside the cirrus added to it that
increases from 0 to $\Delta A_\mathrm{g_\mathrm{SDSS}}=0.194$.

The extinctions found for all cirri and filters are presented in Table 1. It should be emphasized that these are the estimates of the extinctions in the cirri themselves that do not 
include the additional background extinction in the comparison regions, which is worth adding to the extinction specified in Table 1 if it is necessary to estimate the total 
extinction through the entire dust in the directions to these cirri. We estimated the uncertainties in the extinctions based on the number of stars being used, their distribution
in absolute magnitude based on PARSEC and TRILEGAL, the fluctuations of this distribution, the photometric accuracy, and the uncertainty in the difference of the positions of the 
cirrus and the comparison region in space. The latter uncertainty is very significant (at least $\pm0.02$ mag) due to the corresponding uncertainty in the distances to the cirri 
and the comparison regions, although the obvious proximity of the cirri under consideration to the Sun makes this uncertainty acceptable. The results obtained are reliable 
statistically, since thousands of stars were used in each region under consideration (cirrus or comparison region).

To check the results obtained, for each cirrus we compared the reddening $E(g_\mathrm{SDSS}-r_\mathrm{SMSS})=A_\mathrm{g_\mathrm{SDSS}}-A_\mathrm{r_\mathrm{SDSS}}$ calculated from 
Table 1 with the same reddening found by a completely different method: as the difference in themodes of the distribution of red dwarfs in $g_\mathrm{SDSS}-r_\mathrm{SMSS}$ color 
in the cirrus and the comparison region. The reddenings derived by these two methods agree within the uncertainty limits.

For each cirrus Table 3 gives four estimates of the distance range in which the cirrus is located found in our study (averaged over all filters) and from the G17,
GSZ19, and LVB22 reddening maps. The ranges specified in Table 3 also include their uncertainties. The uncertainty in the distribution of the stars being
used on absolute magnitude dominates in these uncertainties.

It can be seen from Table 3 that all four distance estimates overlap significantly only for cirri 2 and 5, i.e., the largest ones and with the greatest extinction. The uncertainty 
in the distances to the remaining cirri is apparently caused precisely by their smaller size and extinction. The range of overlap between the distance estimates $R^{*}$ may be 
considered as the most probable estimate of the distance to the center of the corresponding cirrus. This range is specified in the lower row of Table 3. For cirrus 4 there are
no overlaps of the ranges for all four estimates, but we specified the overlap of the ranges for it without taking into account the estimate from LVB22, which differs greatly 
from the remaining ones. Note that for cirrus 4 we revealed a slight increase in extinction at $R>500$ pc. Cirrus 4 may be a projection of two cirri at different distances from 
the Sun onto the celestial sphere. For such a high-latitude cirrus thismeans the existence of dust at a distance $|Z|>445$ pc from the Galactic midplane. This increase in extinction 
may also be the result of some error because only a small part of cirrus 4 is inside Stripe 82.

To determine the observed extinction law in each cirrus, the extinction estimates from Table 1 are represented by the black diamonds in Fig. 4 as a function of the reciprocal of 
the wavelength $1/\lambda_\mathrm{eff}$ in micrometers. It can be seen that for all cirri the linear fit to the extinction law (with its own coefficient for each cirrus) is better 
than the fit by the CCM89 extinction law with any $R_\mathrm{V}$. This is particularly clearly seen for all cirri in the NUV filter and for cirrus 1 in the $z_\mathrm{SDSS}$ and 
$i_\mathrm{SDSS}$ filters, where the CCM89 extinction law with any $R_\mathrm{V}$ cannot be drawn through the extinctions found. Given the uncertainties, our results confidently
show a very slight increase in extinction with $1/\lambda_\mathrm{eff}$ for cirrus 1 and do not rule out a comparatively slight increase for the remaining cirri. This suggests a
significant contribution of the gray extinction to the total extinction.

Nevertheless, for a qualitative estimate of the observed extinction law in comparison with CCM89 with $R_\mathrm{V}=3.1$, we calculated the optimal $R_\mathrm{V}$ and the CCM89 law 
with it is indicated by the red curve in Fig. 4. By the optimal $R_\mathrm{V}$ we mean the one that is closest to 3.1, but lets the extinction law to pass through our extinction 
estimates for the shortwavelength filters within the limits of their uncertainties and maximally close to our estimates for the long-wavelength filters. The optimal coefficients are
$R_\mathrm{V}=6.9$, 4.0, 5.1, 2.9, and 3.7 for the five cirri, respectively. Since these numbers are not full-fledged statistical estimates and the extinction law with the
optimal $R_\mathrm{V}$, obviously, is worse than the linear fit, we did not estimate the uncertainties in these optimal coefficients. For all of the cirri, except the fourth one,
the optimal $R_\mathrm{V}>3.1$. This agrees qualitatively with the estimates from the map of spatial $R_\mathrm{V}$ variations obtained by Gontcharov (2012): $3.4<R_\mathrm{V}<3.9$ for
the cirri under consideration if the distances from Table 3 are adopted for them. However, according to Gontcharov (2012), at high latitudes $R_\mathrm{V}$ depends significantly on 
distance. And the low value of the optimal $R_\mathrm{V}$ for cirrus 4 can be explained by the component of this cirrus far from the Sun found by us, since, according to 
Gontcharov (2012), $R_\mathrm{V}<3.1$ at $R>500$ pc.

A slight increase in extinction with $1/\lambda_\mathrm{eff}$, i.e., a manifestation of gray extinction, along with significant spatial variations of the extinction law, are typical
for the dust medium far from the Galactic midplane (at a distance $|Z|>200$ pc from it), as shown by Gorbikov and Brosch (2010), Davenport et al. (2014), and 
Gontcharov (2012, 2013, 2016a, 2017b).

It is worth comparing our results for cirrus 2 with the results of Szomoru and Guhathakurta (1999), who investigated another part of this cirrus (designated by them as the PV1 cloud) 
offset by $2\circ$ from Stripe 82 using Wolf diagrams. The best CCM89 extinction law with a very small $R_\mathrm{V}\approx1.7$ found by them for this part of the cirrus deserves 
special attention. This seemingly contradicts the weakly growing extinction law close to CCM89 with $R_\mathrm{V}\approx4$ that was found by us for cirrus 2. Therefore,
we will point out the differences in these studies. The part of cirrus 2 investigated by Szomoru and Guhathakurta (1999) has approximately twice the extinction $A_\mathrm{V}$ 
and, therefore, may possess different properties, including a different extinction law. The area investigated by them is smaller than the area investigated by us approximately 
by an order of magnitude. The $UBVRI$ photometry used by them is less accurate and less deep than that used by us. As a result, the uncertainties in the extinctions found
by them are approximately a factor of 3 larger than our ones (see their Figs. 4 and 6). And the conclusion of Szomoru and Guhathakurta (1999) about a small $R_\mathrm{V}\approx1.7$
is not reliable enough: their results for different filters are described by the CCM89 law with the wide range $1.5<R_\mathrm{V}<3.5$ almost equally well. Thus, comparison of our 
results with those of Szomoru and Guhathakurta (1999) primarily shows progress in this field of astronomy over 23 years.

For comparison with our results, in Fig. 4 we presented the estimates from SFD98 with their uncertainties, combined with the CCM89 extinction law at $R_\mathrm{V}=3.1$, as the green 
regions between the effective wavelengths of the $B$ and $V$ filters indicated by the vertical straight lines. It can be seen that for all cirri the SFD98 estimates agree with our 
extinction estimates in the $g_\mathrm{SDSS}$ and $r_\mathrm{SMSS}$ filters, given their uncertainties (two black diamonds are inside or touch the green regions by the uncertainty bars). 
This means that our results validate the IR emission--reddening calibration for SFD98 in the wavelength range between the $B$ and $V$ filters. In particular, this means that the 
estimates of the extinction inside the cirri $\Delta A_\mathrm{V\,SFD98}$ from Table 2 are plausible. However, it follows from Fig. 4 that the extrapolation of the SFD98 estimates 
(green regions) leftward and rightward will encompass our results for other filters, perhaps, only for cirrus 4. Consequently, if our estimates of the extinction law are valid, 
then, in general, it is not worth extrapolating the SFD98 predictions to other wavelength ranges. This conclusion is particularly important for cosmology and research on
extragalactic objects, as shown by Bogomazov and Tutukov (2011).

\section*{DISCUSSION OF THE EXTINCTION LAW}

Chilingarian et al. (2010), Chilingarian and Zolotukhin (2012), and Chilingarian et al. (2017, hereafter CZK17) showed that the elliptical galaxies without current star formation 
exhibiting a low redshift, once their color was dereddened and the k-correction was applied, form a compact `red sequence' with a comparatively low color dispersion on the 
color--absolute magnitude diagrams. This allows the observed extinction laws in the cirri and the comparison regions under consideration to be independently compared by
analyzing the colors of such galaxies.

To select elliptical galaxies in the cirri and the comparison regions under consideration, we used the sample of Bottrell et al. (2019), which contains the results of a photometric 
decomposition of almost 17000 galaxies in SDSS Stripe 82 with measured spectroscopic redshifts. We could not reproduce the galaxy selection criteria applied in CZK17, since
in the cirri and the comparison regions under consideration only a minority of elliptical galaxies have accurate photometric estimates in the GALEX NUV band that underlie the 
CZK17 criteria. However, we obtained a sufficiently complete sample of redsequence elliptical galaxies using other criteria: the Se\`ersic index $n>3.5$, the apparent ellipticity
$e<0.5$, and the contribution of the disk component to the total luminosity $D/T=0$. It is worth noting that our sample includes galaxies with photometric profiles close to the 
de~Vaucouleurs ones with $n=4$ (de Vaucouleurs 1948) or even steeper, which is typical for bright elliptical galaxies (Caon et al. 1993). From the produced sample we selected only 
those galaxies for which the errors of the photometry in both $g$ and $r$ bands were less than 0.1 (we used the Petrosian magnitudes following CZK17).

There were about 200 galaxies in each cirrus and comparison region. The redshift of these galaxies is comparatively low (it does not exceed $z=0.3$ and has a median $z=0.1$), 
which is needed for a high accuracy of calculating the k-correction. The distribution of galaxies in absolute magnitude is approximately the same in the cirri and the corresponding 
comparison regions and agrees with the distribution of the galaxies used in CZK17 at the same redshifts. We corrected the colors and absolute magnitudes of the galaxies for 
reddening and extinction based on the SFD98 map using the CCM89 extinction law with $R_\mathrm{V}=3.1$. In addition, we applied the k-correction with the k-corrections 
calculator\footnote{http://kcor.sai.msu.ru/} using analytical fits from Chilingarian et al. (2010).

The colors of the galaxies corrected by us must differ systematically in the cirri and the comparison regions under consideration if the extinction laws are different in them. 
For a flatter extinction law in the cirrus than that in the corresponding comparison region, i.e., at a comparatively higher extinction in the long-wavelength part of the spectrum, 
as in Fig. 4 for cirrus 1, the colors of the galaxies will be corrected for an overestimated reddening and will be bluer. This effect must be more pronounced for a larger wavelength
difference of the pair of filters under consideration, i.e., for example, for the $u_\mathrm{SMSS}-z_\mathrm{SMSS}$ color. However, the photometry in the uSDSS band turned out to be 
insufficiently accurate. The most promising of the remaining colors, $g_\mathrm{SDSS}-i_\mathrm{SDSS}$ and $g_\mathrm{SDSS}-z_\mathrm{SDSS}$, showed very similar results. 
Therefore, we discuss only the $g_\mathrm{SDSS}-z_\mathrm{SDSS}$ color variations. The median of the $g_\mathrm{SDSS}-z_\mathrm{SDSS}$ color for each cirrus and comparison region 
is presented in Table 4. As the uncertainty in the color found we specified the standard deviation of the mean following the CZK17 approach.

Table 4 shows that only in cirrus 1 the galaxies are noticeably (by $\Delta(g_\mathrm{SDSS}-z_\mathrm{SDSS})=0.050\pm0.014$) bluer than those in the corresponding comparison
region. On the other hand, Fig. 4 shows that, according to the Wolf diagrams, the stars in cirrus 1 are bluer than those according to the CCM89 law with 
$R_\mathrm{V}=3.1$ by $\Delta(g_\mathrm{SDSS}-z_\mathrm{SDSS})=0.13\pm0.05$. The results from the stars (Wolf diagrams) and from the colors of galaxies can differ primarily due
to the difference in the distribution of the stars and galaxies being used in the sky in combination with our fairly rough identification of the cirri as $6.1\times6.1$
cells in accordance with the angular resolution of the SFD98 map. We attributed or did not attribute these cells to a cirrus entirely. However, since the filaments of the cirri can 
be thinner than 6.1 arcmin,\footnote{The structure of the cirri is discussed by Marchuk et al. (2021). Panopoulou et al. (2022) showed that the characteristic thickness of a 
molecular cloud filament is 0.1 pc, implying 1.5--2 arcmin for the cirri considered by us. Panopoulou et al. (2022) used meaningful data, but made a methodological error that they 
corrected in the Corrigendum to their paper.} 
only part of such a cell may contain a cirrus. Consequently, only some of the galaxies in the cirrus cell actually serve for it as a background and demonstrates the same extinction
law. In such a situation the galaxies in the cirrus cell that are actually outside its projection onto the sky increase the estimate of $g_\mathrm{SDSS}-z_\mathrm{SDSS}$ for the cirrus.
Similarly, a decrease in the estimate of $g_\mathrm{SDSS}-z_\mathrm{SDSS}$ in the comparison region is also possible.

In addition, the disagreement in the results for the stars and galaxies can be explained by a change of the extinction law with distance if there are three dust layers toward the 
cirrus (ahead of, inside, and behind the cirrus) with different extinction laws. The first layer gives the extinction and the extinction law for the comparison region, the first 
and second layers give those for the cirrus, and all three layers give those for the galaxies. As noted previously, such a change of the extinction law with distance at high 
latitudes was found by Gontcharov (2012).

Note that in all of the cirri and the comparison regions the galaxies turned out to be slightly bluer than the color predicted by CZK17 for the red sequence using their system 
of equations (1), depending on the absolute magnitude. For all of the cirri and the comparison regions the median of the predicted color turned out to be the same: 
$g_\mathrm{SDSS}-z_\mathrm{SDSS}=1.440\pm0.001$ (the standard deviation of the mean is specified as the uncertainty). The red sequence selected by us turned out to be slightly 
bluer than the predicted one apparently due to the differences in the selection criteria.

\section*{CONCLUSIONS}

We used the star counts on Wolf diagrams to determine the interstellar extinction in five Galactic cirri (filamentary dust clouds with a low extinction) in SDSS Stripe 82, at middle 
and high Galactic latitudes. For this purpose, we used UV photometry for stars in the GALEX NUV filter and deep photometry for red dwarfs in five SDSS 
$u_\mathrm{SDSS}$, $g_\mathrm{SDSS}$, $r_\mathrm{SDSS}$, $i_\mathrm{SDSS}$, $z_\mathrm{SDSS}$ bands and four SMSS $g_\mathrm{SMSS}$, $r_\mathrm{SMSS}$, $i_\mathrm{SMSS}$, 
$z_\mathrm{SMSS}$ bands. We identified the cirri as sky regions with an enhanced IR emission from the SFD98 map. The extinction in them was calculated relative to the nearby 
comparison regions with a reduced IR emission according to SFD98. The results for different filters agree well, giving the range of distances and the extinction law for each cirrus. 
The distances in the range 140--415 pc found agree well with the 3D reddening maps of Gontcharov (2017a), Green et al. (2019), and Lallement et al. (2022). 

In the wavelength range between the $B$ and $V$ filters the extinctions found are consistent with the estimates from SFD98 for the CCM89 extinction law with $R_\mathrm{V}=3.1$. 
However, for the shorter- and longerwavelength filters the extinctions found deviate significantly from this law. The observed extinction law is best described as the inverse 
proportionality of the extinction and wavelength with its own coefficient for each cirrus. Our results confidently show a very slight decrease in extinction with wavelength for 
cirrus 1 and do not rule out a comparatively slight decrease for the remaining cirri. This suggests a significant contribution of the gray extinction to the total extinction. 
These results are consistent with the previous measurements of the extinction law far from the Galactic midplane presented in the review of Gontcharov (2017b). 

This study may be considered as a successful pilot test of the Wolf diagram method on present day observational data. In future, one can invoke photometric data from deep surveys 
in other filters and apply a more refined initial identification of cirrus filaments.

\section*{ACKNOWLEDGMENTS}

This work was supported by the Russian Science Foundation (project no. 20-72-10052). We thank M. Khovrichev for his help in accessing the Internet resources, T. Rakhmatulina and 
S. Antonov for their help in accessing the SkyMapper Southern Sky Survey data, and the referees for their useful remarks. 

In this study we used resources from the Strasbourg Astronomical Data Center (http://cdsweb.ustrasbg.fr), including the SIMBAD database and the X-Match service. This study uses the 
Filtergraph online data visualization system (Burger et al. 2013, https://filtergraph.com). This study is based on data from the Sloan Digital Sky Survey (http://www.sdss3.org/). 
We used digital products from the SkyMapper Southern Sky Survey (https://skymapper.anu.edu.au), which belongs to and is operated by the Research School of Astronomy and Astrophysics 
at the Australian National University. The study uses the K-corrections calculator accessible at http://kcor.sai.msu.ru/ and data from Galaxy Evolution Explorer, GALEX, one of the 
NASA missions operated by the Jet Propulsion Laboratory.

\section*{REFERENCES}


\newpage

\end{document}